# Rescaling of spatio-temporal sensing in eukaryotic chemotaxis

Keita Kamino, [1,¶,*] Yohei Kondo, [2,¶,*]

[1]FOM Institute AMOLF, Amsterdam, Netherlands, [2]Graduate school of Informatics, Kyoto University, Kyoto, Japan,

*Corresponding author

Email:

k.kamino@amolf.nl (KK)

ykondo@sys.i.kyoto-u.ac.jp (YK)

[¶]These authors contributed equally to this work.

# Abstract

Eukaryotic cells respond to a chemoattractant gradient by forming intracellular gradients of signaling molecules that reflect the extracellular chemical gradient – an ability called directional sensing. Quantitative experiments have revealed two characteristic input-output relations of the system: First, in a static chemoattractant gradient, the shapes of the intracellular gradients of the signaling molecules are determined by the relative steepness, rather than the absolute concentration, of the chemoattractant gradient along the cell body. Second, upon a spatially homogeneous temporal increase in the input stimulus, the intracellular signaling molecules are transiently activated such that the response magnitudes are dependent on fold changes of the stimulus, not on absolute levels. However, the underlying mechanism that endows the system with these response properties remains elusive. Here, by adopting a widely used modeling framework of directional sensing, local excitation and global inhibition (LEGI), we propose a hypothesis that the two rescaling behaviors stem from a single design principle, namely, invariance of the governing equations to a scale transformation of the input level. Analyses of the LEGI-based model reveal that the invariance can be divided into two parts, each of which is responsible for the respective response properties. Our hypothesis leads to an experimentally testable prediction that a system with the invariance detects relative steepness even in dynamic gradient stimuli as well as in static gradients. Furthermore, we show that the relation between the response properties and the scale invariance is general in that it can be implemented by models with different network topologies.

# Introduction

Many eukaryotic cells exhibit chemotaxis – the ability to sense and move up or down spatial gradients of chemicals. Chemotaxis underlies many biological phenomena such as cancer metastasis, immune response, wound healing and embryonic development [1–3]. In a chemoattractant gradient, cells are constantly monitoring the direction of the gradient by means of chemical reactions on and within the cell membrane, forming gradients of signaling molecules in the cytosol directed toward the extracellular gradient. This process, functioning like an internal chemical compass, is referred to as directional sensing. In spite of intensive molecular genetic study [3], the system-level design principle that governs the flexible and dynamic behavior of gradient sensing has remained elusive.

The signal transduction events in eukaryotic gradient sensing have been most intensively studied in *Dictyostelium* and neutrophils [4]. Although molecular species of chemoattractant can vary between the type of cells, many of identified molecular components of the signaling system are conserved across cell types [3]. A chemotactic response is initiated by binding of chemoattractants to G-protein coupled receptors (GPCRs) on the cell membrane [3–5]. The binding causes dissociation of the G-protein α and βγ subunits within the cell membrane [6,7]. Studies suggest that the βγ subunit mediates the activation of downstream effectors such as Ras proteins [8]. The switching of the activity of Ras proteins is regulated by multiple guanine nucleotide exchange factors (GEFs) and GTPase-activating proteins (GAPs) [9]. GTP-bound Ras proteins activate phosphoinositide 3-kinase (PI3K) that synthesizes phosphatidylinositol 3,4,5-triphosphate ($PIP_3$) from phosphatidylinositol 4,5 –bisphosphate ($PIP_2$). Accumulation of $PIP_3$ on the membrane recruits multiple PH domain-containing proteins to the membranes such as Cytocolic

Regulator of Adenylyl Cyclase (CRAC), protein kinase BA and PH domain protein A (PhdA), which then triggers the force-generating process of actin polymerization, or pseudopod extension [3].

Live cell analyses have revealed spatio-temporal properties of the directional sensing system. The output molecules of the system, such as Ras, PI3K and $PIP_3$, exhibit localization to the site of highest concentration of chemoattractant, forming concentration gradients intracellularly [10–14]. The steepness of the intracellular molecules can be steeper than that of a chemoattractant gradient, i.e., the directional signal is amplified [15]. The outputs are essentially separable from the downstream signaling modules that directly controls the motility of cells because even a cell immobilized by an inhibitor of actin polymerization shows similar localization of the signaling molecules [3,11,14,15]. Importantly, although the localization is persistent as long as a chemoattractant gradient is around, the same intracellular molecules shows only transient accumulation on the membrane upon a spatially homogeneous temporal elevation of the stimulus, i.e., the response shows adaptation [12,16] – an ubiquitous behavior observed across many biological systems [17–19].

A conceptual scheme called local excitation and global inhibition (LEGI) has been proposed to explain the rich behavior of the directional sensing system [11]. The basic idea of the LEGI hypothesis is that the binding of chemoattractant molecules to the receptors on the membrane elicits two counteracting processes, i.e., excitation and inhibition of the output signals. Both excitation and inhibition are persistent as long as the stimulus is around, but they operate in different spatio-temporal scales. The excitation process is fast but has a shorter range of action, so the degree of excitation at each site on the membrane reflects the local chemoattractant concentration. On the other hand, the inhibition is slow but has a longer range of action, and therefore the local level of inhibition, being spatially averaged, reflects the mean chemoattractant concentration over the cell periphery. Upon a spatially uniform stimulus, the fast nature of the excitation causes a rapid increase in the output level, but the slow inhibition process counteracts the excitation process gradually, resulting in adaptation at a steady state. In a spatially-graded stimulus, on the other hand, the degree of excitation differs along the stimulus gradient whereas the level of inhibition is nearly uniform. As a result, excitation exceeds inhibition at the site of higher chemoattractant concentration and inhibition exceeds excitation at the lower side, resulting in a persistent graded response. Although direct experimental evidences for the LEGI mechanism are still limited, many LEGI-based mathematical models have been proposed and explained the behavior of directional sensing at least qualitatively [1,14,16,20–23].

Crucially, experiments on *Dictyostelium* have revealed two characteristic features of gradient sensing that can potentially challenge the LEGI hypothesis. First, by simultaneously visualizing the gradients of the output signals and chemoattractant concentration, Janetopoulos et al have shown that the spatial profile of the output signal is determined by the relative steepness of the chemoattractant gradient instead of its absolute level [15]. Second, by measuring the adaptation dynamics of the output signal upon spatially-uniform step stimuli, Takeda et al have demonstrated that the response magnitude is dictated approximately by fold changes of the input stimuli rather than absolute levels [16,24–26]. Recently, a mathematically idealized response property of the second type has been named fold-change detection (FCD), where the entire response shape depends only on temporal fold changes in input and not on absolute levels [24–28]. Similarly, one can think of an idealized response property to gradient stimuli where the spatial profile of the output signal is strictly determined by the relative steepness of input gradients. To discriminate the two response properties where one refers to relative gradient detection in space and the other in time, we hereafter call them spatial FCD and temporal FCD respectively. In spite of apparent similarities between the two FCDs and its potential importance in understanding eukaryotic directional sensing, it remains unanswered whether and how the two are related with each other.

Here, under the LEGI framework, we address three questions: (1) Can a model based on the LEGI mechanism account for both FCDs at the same time? (2) Does either one of the FCDs automatically imply the other? (3) Does a LEGI model that achieves both FCDs require a specific network topology? As a LEGI-based model capable of both spatial and temporal FCD, we propose a scale-invariant LEGI model whose responses exhibit rescaling properties due to the scale invariance, i.e., symmetry, of the system equations. Analyses of the model reveal that the scale invariance can conceptually be divided into two parts, each of which is responsible for spatial and temporal FCD respectively. As another behavioral consequence of having the scale invariance, we propose spatio-temporal FCD, the ability by which a system detects the relative steepness of a dynamic gradient stimulus as well as of a static gradient. Finally, the FCD properties are shown to be independent of the network topology of a model. Our results suggest that the scale invariance is the underlying design principle of the directional sensing system of eukaryotic cells.

# Models

**Scale-invariant LEGI model**

Here we describe the scale-invariant LEGI model capable of both spatial and temporal FCDs. The model contains two signaling components, a membrane-bound activator A, which is considered to be the output of the system, and its inhibitor B. Following the standard LEGI framework [11,20], we here adopt a feedforward regulatory network, where both components A and B are activated by an input signal, i.e., chemoattractant S (Fig. 1). The inhibitor can be detached from the membrane and diffuse inside the cell with fast kinetics; therefore, the inhibitor levels on the cell membrane are globally coupled through the cytosolic inhibitor level. Taken together, the dynamics of each component's activity at each point on the cell membrane is described as

$$\frac{\partial A}{\partial t} = k_a \frac{S^n}{S^n + (KB)^n} - k_{-a}A,$$
$$\frac{\partial B}{\partial t} = k_b S - k_{-b}B + D(\langle B \rangle - B), \quad (1)$$

where $k_a, k_{-a}, k_b, k_{-b}, n, K,$ and $D$ are constant parameters. $\langle B \rangle$ is spatially averaged inhibitor level over the membrane and thus represents the cytosolic level of inhibitor B. The parameter values used in the simulations in the following sections are $k_a = 5, k_{-a} = 0.5, k_b = 0.5, k_{-b} = 0.25, n = 6, K = 1$, and $D = 0.1$, although the FCD properties explained in detail below are not dependent on the specific parameter values. Note that we adopted a relatively high Hill coefficient ($n = 6$) for the amplification of the input signal. We discuss possible molecular mechanisms for the model equations, especially the one for the nonlinear function, $S^n/(S^n + (KB)^n)$ (see Discussion and Supporting Text S1).

**Fig. 1. Schematic representation of the scale-invariant LEGI model.** The model describes the dynamics of the levels of activator A and inhibitor B upon chemoattractant stimulus S. The arrows and the bar-ended arrow represent excitatory and inhibitory regulations between the signaling components, respectively.

**LEGI models capable of either temporal or spatial FCD but not both**

To demonstrate that temporal and spatial FCDs can be implemented independently with each other, here we describe two simple LEGI-based models partially different from the scale-invariant LEGI model (Eqs. 1). The models contain the same two signaling components as in the scale-invariant LEGI model (Eqs. 1), a membrane-bound activator A and its inhibitor B, and retain the basic network topology of the chemical reaction circuit (Fig. 1). The first model differs from the scale-invariant LEGI model in that inhibitor B on the membrane mediates the binding/unbinding processes of the inhibitor to the membrane, introducing a nonlinear term in the equations:

$$\frac{\partial A}{\partial t} = k_a \frac{S^n}{S^n + (KB)^n} - k_{-a}A,$$
$$\frac{\partial B}{\partial t} = k_b S - k_{-b}B + DB(\langle B \rangle - B), \quad (2)$$

where $k_a, k_{-a}, k_b, k_{-b}, n, K$, and $D$ are constant parameters. The parameter values used in the simulations below are $k_a = 5$, $k_{-a} = 0.5$, $k_b = 0.5$, $k_{-b} = 0.25$, $n = 6$, $K = 1$, $D = 1$. The parameter values of this and the following models (i.e., Eqs. 3 and 4) are chosen so that they show response magnitudes and timescales comparable to the scale-invariant LEGI model (Eqs. 1), but the FCD properties of the models described in this work do not depend on the parameter values.

In the second model, the inhibitor B now directly breaks down the activator A, instead of interfering with the synthesis of it:

$$\frac{\partial A}{\partial t} = k_a S - k_{-a}AB,$$
$$\frac{\partial B}{\partial t} = k_b S - k_{-b}B + D(\langle B \rangle - B), \quad (3)$$

where $k_a, k_{-a}, k_b, k_{-b}$, and $D$ are constant parameters. The parameter values used in the simulations are $k_a = 0.5$, $k_{-a} = 0.5$, $k_b = 1$, $k_{-b} = 1$, and $D = 2$.

**Scale-invariant LEGI model based on a negative feedback loop**

Here we describe another scale-invariant LEGI model to show that the feedforward regulatory network is not essential to realize spatial and temporal FCD. The only difference here is that the interaction between A and B follows what is called a negative feedback loop. The equations are

$$\frac{\partial A}{\partial t} = k_a \frac{S^n}{S^n + (KB)^n} - k_{-a}A,$$
$$\frac{\partial B}{\partial t} = k_b AB - k_{-b}B + D(\langle B \rangle - B), \qquad (4)$$

where $k_a, k_{-a}, k_b, k_{-b}, n, K$, and $D$ are constant parameters. The parameter values used in the simulations are $k_a = 0.5$, $k_{-a} = 0.5$, $k_b = 2.5$, $k_{-b} = 0.25$, $n = 6$, $K = 1$, and $D = 0.5$,

**Numerical method**

In numerical computation, we considered circular cell periphery, and discretized the periphery into $N =$ 100 points. We utilized MATLAB's "ode45" function for numerical integration of the models.

# Results

**Responses to spatially-homogeneous time-varying inputs**

First we analyze the response of the scale-invariant LEGI model (Eqs. 1; see Models) based on an incoherent feedforward loop (iFFL) (Fig. 1) to spatially-uniform stimuli. Without gradient stimuli, the model equations become

$$\frac{dA}{dt} = k_a \frac{S^n}{S^n + (KB)^n} - k_{-a}A,$$
$$\frac{dB}{dt} = k_b S - k_{-b}B, \qquad (5)$$

because the average inhibitor level $B$ along the cell membrane equals to that at each point, i.e., $\langle B \rangle = B$. Thus, the system is now characterized by three scalar variables $A$, $B$ and $S$. When exposed to a step input $S$, the activator level $A$ increases transiently and adapts to a prestimulus level, whereas the inhibitor level $B$ monotonically increases until it reaches a steady-state value (Fig. 2a). The steady-state concentration of each variable is given by

$$A = \frac{k_a}{k_{-a}} \frac{k_{-b}^n}{k_{-b}^n + (Kk_b)^n}, \quad B = \frac{k_b}{k_{-b}} S_0, \qquad (6)$$

where $S_0$ represents the level of the input stimulus $S$ at a steady state. The steady-state level of the activator $A$ is independent of the stimulus level $S_0$, meaning that the system shows perfect adaptation. Furthermore, the system responds to fold changes in input but not its absolute level, as demonstrated by the responses to two successive step inputs with identical fold change (Fig. 2b). Shoval and co-workers have shown that temporal FCD results from invariance of system equations upon a scale transformation of variables [27]. In our model, temporal FCD is attributed to the invariance of the equations (Eqs. 5) under the scale transformation

$$(A, B, S) \rightarrow (A, pB, pS) \ (p > 0), \qquad (7).$$

Note that the activator $A$ remains invariant when other variables are multiplied by a scalar $p$. Intuitively, the invariance means that multiplying an input stimulus $S$ by $p$ does not affect the trajectory of $A$, whereas the trajectory of $B$ is multiplied by $p$.

**Fig. 2. Temporal fold-change detection (FCD) of the scale-invariant LEGI model in response to spatially uniform stimuli.** The insets show the input variable $S$ (green). (a) Trajectories of the activator level $A$ (red) and the inhibitor level $B$ (blue) upon single step input. (b) A trajectory of the activator level $A$ upon two-step inputs with identical fold change ($p = 5$) but different absolute levels.

**Responses to static spatially-graded inputs**

Next we focus on responses to static spatially-graded stimuli. To this end, we simulated responses to linear gradient stimuli in two-dimensional space assuming that the reactions occur on a circular cell membrane. In Figure 3a, the levels of activator $A$ and inhibitor $B$ at points on the membrane are plotted as a function of input level $S$ at the corresponding points. While the inhibitor $B$ is distributed more homogeneously than that of the input stimulus $S$, the activator $A$ shows a steeper gradient than the input stimulus $S$; this means that the gradient signal is amplified. Figure 3b shows the level of activator $A$ and inhibitor $B$ at the front and back of the cell as a function of the absolute input level at the center of the cell $S_0$ with a fixed relative gradient steepness. While the level of inhibitor $B$ increases with the stimulus level $S_0$, the output $A$ remains constant both at the front and back. Thus, the system's output is not dependent on the absolute input level but on the relative steepness of the input, showing spatial FCD. The origin of the spatial FCD property can be understood as follows. Suppose that the system is in the steady state, i.e, $\frac{\partial A}{\partial t} = \frac{\partial B}{\partial t} = 0$, under a gradient stimulus. Then, the right-hand sides of the model equations (Eqs. 1) are invariant under a scale transformation, $(A, B, S) \rightarrow (A, pB, pS)$. That is, the $p$-fold change in the input level $S$ can be perfectly compensated by the multiplied $pB$, and thus the profile of the activator $A$ does not change. Note that here we do not assume that the gradient stimulus is linear, and therefore the system shows spatial FCD in an arbitrary spatial input profile.

**Fig. 3. Spatial FCD of the scale-invariant LEGI model upon static gradient stimuli.** The input profile follows $S(x,y) = S_0 + \Delta S\,x$, where the origin of the spatial coordinate $x$ is located at the cell center. (a) Amplified response in the activator level $A$. The activator level $A$ (red circle) and inhibitor level $B$ (blue circle) are plotted against the input level $S$ at each point on the cell periphery with both $A$ and $B$ normalized by the value at the center of the cell. The input stimulus used is $S_0 = 0.1$ and $\Delta S = 0.01$. The black line indicates normalized $S$, which corresponds to an unamplified response. (b) The activator level $A$ (red) and inhibitor level $B$ (blue) at the cell front (solid line) and back (broken line) are plotted as a function of the mid-point input level $S_0$. The relative gradient is fixed as $\Delta S = 0.1 S_0$.

To illustrate the consequence of the scale invariance, we study the model (Eqs. 1) in one-dimensional geometry where the front and back membrane consisting of single points are connected by cytosol. Here a spatially-graded stimulus is represented by two scalars representing stimulus levels at the cell front $S_f$ and at the back $S_b$ with $S_f$ larger than $S_b$. Thus, the model equations at the front are described as

$$\frac{dA_f}{dt} = k_a \frac{S_f^n}{S_f^n + (KB_f)^n} - k_{-a}A_f,$$
$$\frac{dB_f}{dt} = k_b S_f - k_{-b}B_f + D(\frac{B_f + B_b}{2} - B_f), \quad (8)$$

and the model equations at the back are given by the same expression by exchanging the subscripts $f$ and $b$. The analytical steady-state solutions for the concentration of the activator $A$ and the inhibitor $B$ are written as

$$A_f = \left(\frac{k_a}{k_{-a}}\right) \frac{(S_f/\bar{S})^n}{(S_f/\bar{S})^n + K^n (B_f/\bar{S})^n},$$
$$\frac{B_f}{\bar{S}} = \frac{k_b}{k_{-b}} + \frac{k_b}{2(k_{-b} + D)} \frac{(S_f - S_b)}{\bar{S}}, \quad (9)$$

where $\bar{S} = (S_f + S_b)/2$. The steady-state solutions at the back are given by the same expression by exchanging the subscripts $f$ and $b$. Note that $A_f$ and $A_b$ are dependent only on the relative level of the input stimuli $S_f/\bar{S}$ and $S_b/\bar{S}$, respectively, i.e., the system shows spatial FCD. Therefore, the scale invariance in the model equations results in the normalization of the input stimulus by its spatially-averaged value in the solution.

**Responses to complex input stimuli**

Several experimental studies have investigated responses to more complex input stimuli such as a rapid reversal of the direction of chemoattractant gradients [12], stimulation with multiple sources of chemoattractant [15] and a combination of gradient and uniform stimuli [6]. To test whether our model (Eqs. 1) can reproduce responses to such complex stimuli, we simulated responses following the stimulation protocols of those experiments. First, experiments have demonstrated that the direction of the gradient of the signaling molecules switches promptly upon a reversal of the direction of chemoattractant gradient [12]. The behavior is thought to be incompatible with a model based on the spatial instability in reaction-diffusion system, or "Turing-type pattern formation", whereas LEGI-based models reported so far can often reproduce it [29]. Figure 4a shows the level of activator $A$ at the front and back as a function of time, where the direction of a linear-gradient stimulus is reversed at time $t = 20$. Consistent with the experimental result, our model switches the gradient of the activator $A$. Second, other experiments have shown that cells, when stimulated with multiple sources of chemoattractant, show localizations of the signaling molecule at multiple sites on the membrane in a way that the degree of localization dynamically follows the strength of the stimuli [15]. To confirm the response property of the model, we simulated the response by locating a single point source of stimulus in close proximity to the cell membrane, and then by another point source at the opposite site with increasing level of stimulus (Fig. 4b). In consistent with the experimental results, the model showed multiple peaks of activator $A$ dynamically following the strength of the input stimuli. Third, another interesting property of the gradient sensing system is its 'inversed' responsiveness to a spatially-uniform stimulus: When cells are exposed to a gradient stimulus, then no stimulus (by withdrawing the gradient stimulus), and then a spatially-uniform stimulus, the back of the cell in the gradient stimulus responds more strongly than the front [6]. In Figure 4c, at the initial state the cell is exposed to a linear gradient, then the stimulus is withdrawn, and then a uniform stimulation is applied. In agreement with the experiment, the cell showed higher response at the back upon exposure to the uniform stimulus. These results would be reproducible by a broad range of LEGI-based models, since the response properties rely on the basic machinery of the system, i.e., the difference in the range of action between the activator and inhibitor. Therefore, the results demonstrate that the scale-invariant LEGI model preserves the properties inherent to a LEGI-based model.

**Fig. 4. Responses of the scale-invariant LEGI model to complex stimuli.** (a) The trajectories of the activator $A$ at the cell front (solid line) and back (dotted line) are plotted upon a sudden reversal of a linear gradient ($S(x, y) = S_0 + \Delta S\,x$, $S_0 = 1$ and $\Delta S = \pm 0.1$) at $t = 20$. The inset shows spatial profile of the activator level $A$ at $t = 0$. (b) Responses to two point sources of input stimuli where the input profile is defined as $S(x, y) = C_1 \exp[-((x + x_0)^2 + y^2)/2V] + C_2 \exp[-((x - x_0)^2 + y^2)/2V]$ ($x_0 = 7.5$ and $V = 4$). The positions of the point sources are indicated by the purple squares. (Top left) Spatial profile of the activator level $A$ at the steady state in one point source stimulus, i.e., $C_1 = 1$ and $C_2 = 0$. (Bottom left) Spatial profile of the activator level $A$ at the steady state in two point source stimuli, i.e., $C_1 = C_2 = 1$. (Right) Time series of the activator $A$ at the cell front (solid line) and back (dashed line) upon application of another point source of $S$ at t = 0 at the other side of the cell. The strength of the second signal $C_2$ is dynamically controlled as shown in the inset. (c) Responses of the model upon sequential changes of the input spatio-temporal profile, where a spatially graded stimulus ($S(x, y) = S_0 + \Delta S\,x$, $S_0 = 1$ and $\Delta S = 0.15$) (top left), then no stimulus (top middle) and then a spatially-homogeneous stimulus ($S_0 = 0.3$) (top right) are applied consecutively. Spatial profiles of the activator $A$ at corresponding time points are shown at the bottom.

**Responses to spatially and temporally graded stimuli**

We have shown that our model reproduces both temporal and spatial FCD properties. However, as is suggested by the analyses so far, the scale-invariance conditions that lead to each property overlap only partially and are not exactly the same (Fig. 5a): To achieve the temporal FCD property, the system equations have to be invariant upon the scale transformation (Eq. 7) except the coupling term representing the interaction between different membrane parts, which in our model corresponds to $D(\langle B \rangle - B)$ (Eqs. 1). This is because the coupling term is negligible in a spatially uniform environment where $\langle B \rangle = B$. Let us call this type of invariance temporal FCD symmetry (Fig. 5a). On the other hand, to achieve the spatial FCD property, the equations have to be invariant except the time derivative terms of the variables. This is because the spatial FCD property refers to a steady-state behavior where $\frac{\partial A}{\partial t} = \frac{\partial B}{\partial t} = 0$. We call this type of invariance spatial FCD symmetry (Fig. 5a). Thus, neither the temporal nor spatial FCD property can automatically imply the presence of the other in the LEGI framework.

**Fig. 5. Two types of symmetry in the scale-invariant LEGI model.** (a) Temporal FCD symmetry refers to the scale invariance of the equations excluding the coupling term. Spatial FCD symmetry means the scale invariance of the equations excluding the time derivative terms. (b) A trajectory of the activator level $A$ of the model only with spatial FCD symmetry (Eqs. 2; see Models) upon two-step inputs with identical fold change ($p = 5$) but different absolute levels. The inset shows the input variable $S$. (c) The activator level $A$ (red) and inhibitor level $B$ (blue) at the cell front (solid line) and back (broken line) are plotted as a function of the mid-point input level $S_0$ for the model only with temporal FCD symmetry. The relative gradient is fixed as $\Delta S = 0.1\, S_0$. (d, e) For the model only with spatial FCD symmetry (Eqs. 3; see Models), the responses are plotted as in (b, c).

The above discussion suggests that there exist LEGI-based models that are capable of either temporal or spatial FCD but not both. To demonstrate this, we first consider a model only with temporal FCD symmetry (Eqs. 2; see Models). Here, the equations have the coupling term that includes a nonlinear effect. Numerical simulation shows that the model is capable of temporal FCD but not spatial FCD (Fig. 5b and c), consistent with the observation that the system only has temporal FCD symmetry. Next, we consider another LEGI-based model only with spatial FCD symmetry (Eqs. 3; see Models), where the inhibitor B directly breaks down the activator A instead of interfering with the synthesis of A as in the original model. In contrast to the first model (Eqs. 2), the model shows spatial FCD but not temporal FCD (Fig. 5d and e), which can similarly be understood by the presence of spatial FCD symmetry and the absence of temporal FCD symmetry in the equations. Note that the ways to break each symmetry in the equations are not restricted to the specific functional forms in our example (Eqs. 2 and Eqs. 3); we have chosen arbitrary functional forms for the purpose of demonstration.

The mutual independency of spatial and temporal FCD raises a new question of what is the unique consequence of having both symmetries in one system. We have shown that exposing a system to stimuli changing either in space or time, but not both, can only reveal the presence of either of the symmetries within the system, i.e., spatial or temporal FCD symmetry. Here we demonstrate that input stimuli changing in both space and time at the same time can reveal the simultaneous presence of both symmetries in the system. To this end, we simulated responses of the models to two propagating Gaussian waves with the same shape but different absolute levels (Fig. 6a and b). We found that the scale-invariant LEGI model exhibits exactly the same responses to two wave stimuli if the two inputs are proportional to each other, $S_1(r,t) = pS_2(r,t)$ (Fig. 6c). We hereafter call the property spatio-temporal FCD. Specifically, the activator $A$ shows identical spatio-temporal profile in the two input stimuli (Fig. 6c) whereas the inhibitor $B$ depends on the absolute input level (Fig. 6d). The invariant response is observed irrespective of the speed of the wave (Fig. 6e). In contrast, the models only with partial symmetry (Eqs. 2 and Eqs. 3) cannot show such behavior, showing deviation from spatio-temporal FCD (Fig. 6f and 6g). More precisely, the model only with temporal FCD symmetry shows significant dependence on the absolute input level $S$ when the wave speed is low (Fig. 6f, top panels), whereas the model only with spatial FCD symmetry shows a non-rescaling behavior when the wave speed is high (Fig. 6f, bottom panels). These phenomena can be understood in the following way. If the wave speed is high enough, a traveling wave essentially becomes a short, pulsatile, spatially-uniform stimulus to the cells. Hence, a model with temporal FCD symmetry approximately shows rescaled responses to such stimuli. Similarly, if the wave speed is sufficiently low, a traveling wave becomes more similar to a static gradient stimulus. Under such stimuli, spatial FCD symmetry approximately achieves rescaled responses. Taken together, these results demonstrate that rescaled responses to wave stimuli with various speeds manifest the simultaneous presence of the two symmetries in the system.

**Fig. 6. Spatio-temporal FCD under wave stimuli.** The spatio-temporal profile of the traveling wave stimulus is defined by the Gaussian shape as $S(x, y, t) = s_0 \exp[(x - x_0 - w t)^2 / 2V] + s_b$. Below, $S_{\text{slow}}(x, y, t)$ is defined with the wave velocity $w = 0.5$, while $S_{\text{fast}}(x, y, t)$ is defined with $w = 5$. The width of the wave $V$, the initial position $x_0$, and the basal stimulus level $s_b$ are fixed at $V = 100$, $x_0 = 50$, and $s_b = 0.1$, respectively, in both wave stimuli. We use $p = 10$ for the fold-change parameter. (a) Schematic representation of the numerical experiment where a wave stimulus is applied to a cell. (b) The input level $S$ at the cell front (solid line) and back (broken line) under $S_{\text{slow}}(x, y, t)$. (c) Trajectories of the activator level $A$ at the cell front (solid line) and back (broken line) under $S_{\text{slow}}$ (red) and $pS_{\text{slow}}$ (blue) are plotted for the scale-invariant model (Eqs. 1; see Models). (d) Trajectories of the inhibitor level $B$ are plotted as in (c). (e) For the scale-invariant LEGI model (Eqs. 1), spatio-temporal profiles of the activator level $A$ along the cell periphery are plotted under wave stimuli $S_{\text{slow}}$ (top left), $pS_{\text{slow}}$ (top right), $S_{\text{fast}}$ (bottom left), and $pS_{\text{fast}}$ (bottom right), respectively. Note that each point at the cell periphery is indicated by an angle expressed in radians from 0 (cell front) to $\pi$ (cell back). (f) Responses of the model only with temporal FCD symmetry (Eqs. 2; see Models) are plotted as in (e). (g) Responses of the model only with spatial FCD symmetry (Eqs. 3; see Models) are plotted as in (e).

**Scale-invariant LEGI model with a different network topology**

The interaction between signaling components in the scale-invariant LEGI model (Eqs. 1) follows an iFFL network topology (Fig 1), where the stimulus S activates both the activator A and the inhibitor B (Fig. 1). Although we have explored the scale-invariant properties and their behavioral consequences solely based on the iFFL-based models, it has remained unanswered whether the results depend on the specific network topology. We found that, within the LEGI framework where the activator operates locally and the inhibitor globally, it is also possible to implement the system with a negative feedback interaction (Fig. 7a) that has exactly the same symmetry as the scale-invariant LEGI model (Eqs. 4; see Models). As the result, this model also shows temporal FCD (Fig. 7b), spatial FCD (Fig. 7c) and spatio-temporal FCD (Fig. 7d). Although the detailed feature of the response, such as the overshoot response in the negative feedback model (Fig. 7b), can be specific to a type of network topology, the results clearly demonstrate the FCD properties can be implemented without using an iFFL network toplogy.

**Fig. 7. Responses of the scale-invariant LEGI model with a feedback network topology.** (a) A schematic representation of the regulatory network of the model (Eqs. 4; see Models). (b) A trajectory of the activator level $A$ upon two-step inputs with identical fold change ($p = 5$) but different absolute levels.

The inset shows the input variable *S*. (c) The activator *A* (red) and inhibitor *B* (blue) at the cell front (solid line) and back (broken line) under linear profiles of gradient, $S(x,y) = S_0 + \Delta S\, x$, are plotted as a function of the mid-point input level $S_0$. The relative gradient is fixed as $\Delta S = 0.1\, S_0$. (d) Responses to wave stimuli are visualized as in Figures 6e. Specifically, spatio-temporal profiles of the activator level *A* along the cell periphery under wave stimuli $S_{\text{slow}}$ (top left), $pS_{\text{slow}}$ (top right), $S_{\text{fast}}$ (bottom left) and $pS_{\text{fast}}$ (bottom right), respectively.

# Discussion

Quantitative characterizations of input-output relations of a signaling system place constraints on possible mechanisms. Inspired by two characteristic features of the eukaryotic directional sensing system, i.e., the rescaling responses to temporal changes [16,24] and static spatial gradients of chemoattractant stimuli [15], we asked whether and how temporal and spatial FCD are related with each other under the LEGI framework. By analyzing the scale-invariant LEGI model that is capable of both temporal and spatial FCD, we have identified two types of scale invariance, or symmetry, of the system equations that yield the respective FCD properties (Fig. 5a). It has also been shown that neither of the temporal nor spatial FCD entails the other (Fig. 5). On top of the FCDs, the scale-invariant LEGI model has been shown to reproduce response properties observed in complex input stimuli (Fig. 4). Furthermore, we have demonstrated that the rescaled responses are achievable by a negative feedback model (Fig. 7) as well as an incoherent feed-forward model, a network motif often adopted in the LEGI scheme [1,14,16,20–23]. In spite of this, we note that, in the case of *Dictyostelium* [16], an experimental result support a feed-forward mechanism, raising the possibility that an iFFL mechanism is more advantageous.

Because the LEGI framework by itself does not specify the details of mathematical models, many versions of LEGI-based models have been proposed [1,14,16,20–23,25,26,29]. We have found that some of the preceding models are capable of either temporal or spatial FCD, but not both. Based on the identified mathematical conditions for the two FCDs, we have analyzed some representative models in the literature (see Supporting Text S1). Regarding temporal FCD in the eukaryotic directional sensing system, Adler et al [24] have successfully explained experimental data on rescaled responses in time [16] by using a model with temporal FCD symmetry, although the model have not focused on the spatial aspects of the directional sensing system. On the other hand, several LEGI-based models, as pointed out by Nakajima et al [14], are capable of spatial FCD. For example, Levchenko and Iglesias, in their pioneering work [20], have proposed LEGI-based models capable of spatial FCD. We show in the Supporting Text S1 that some of the models with spatial FCD can be modified to show temporal FCD as well. Note that lack of the symmetries does not necessarily mean that the model is inappropriate or inferior to those with the symmetries. For example, a detailed experimental study has revealed that the activity of Ras, an output signaling molecule of the directional sensing system, upon spatially-uniform stimulation shows slightly imperfect adaptation in *Dictyostelium* [14]. This means that temporal FCD, which entails perfect adaptation [27,28], is only approximately realized and therefore a detailed model would need to break temporal FCD symmetry to make it more realistic. Still, our simplified model has allowed us to gain an understanding of the mathematical relations between the symmetry and the rescaling behavior in the directional sensing system.

Our analysis provides an insight in designing experiments on the directional sensing system. Most experimental studies so far, except a few most recent studies described below, have investigated response properties to input stimuli with either temporal or spatial change but not both. However, as we have demonstrated in figure 6, the presence of the scale-invariant mechanism and resultant spatio-temporal FCD can be directly tested by applying dynamic gradient stimuli to cells. Since the LEGI framework still remains a hypothetical mechanism for gradient sensing, it would be informative to characterize the response properties in dynamic gradient stimuli. We note that such an experiment has become technically possible by using microfluidic systems. For example, recent chemotaxis studies have made use of microfluidic chambers that can generate Gaussian traveling waves [14,30].

Experiments have revealed that the directional sensing system of *Dictyostelium* exhibits highly asymmetric responses at the front and back of propagating Gaussian waves [14,30]: The activated form of Ras shows strong localization toward the side of higher chemoattractant concentration only in the wavefront, where the chemoattractant concentration is increasing in time. In our model, the rescaling response properties are compatible with the asymmetric responsiveness (Fig. 6c). In the preceding works [14,30], the asymmetric behavior has been successfully explained by using LEGI-based models. There, strong amplification of input signal has been crucial for the asymmetric response. Since our scale-invariant LEGI model retains features of a LEGI-based model, it can also attain amplification of a gradient stimulus (Fig. 3a) by adopting comparatively high amplification parameter (i.e., *n* in Eqs. 1). When exposed to a Gaussian traveling wave, this leads to a highly asymmetric response at the front and back of the wave as demonstrated in Fig. 6c.

What is the function of spatio-temporal FCD in eukaryotic chemotaxis? A similar question has been asked about temporal FCD in the context of bacterial chemotaxis [27,31]. When a bacterial cell moves around in an input (i.e., chemoattractant) field, instead of directly monitoring the spatial gradient of a chemoattractant, it gains the information of gradient direction from temporal changes of the input level that it experiences during the incessant random motion [32]. In this setup, it has been shown that temporal FCD ensures that the cell's search, or chemotaxis performance, depends only on the shape of the input field irrespective of the absolute level of it [27]. Relying only on the shape of an input field is considered to be advantageous in searching for a nutritionally favorable environment because of the nature of the physics of the chemical gradient formation [27]. In eukaryotic chemotaxis, the situation is more complicated because the cells sense both temporal and spatial gradient in an input field [14,30]. Therefore, it has not been immediately clear how a cell can achieve a similar chemotaxis property without depending on the absolute input level. Under the assumption that cells' migration pattern is controlled by the output of the directional sensing system, our analyses suggest that cells achieve the chemotaxis property only when they are equipped with both temporal and spatial FCD: As we have demonstrated in figure 6, the behavior of a cell equipped with either of them, but not both, is dependent on the absolute input levels in dynamic chemical gradients. In that sense, spatio-temporal FCD in eukaryotic chemotaxis can be considered as a functional equivalent of temporal FCD in bacterial chemotaxis.

For the sake of simplicity, we have implemented the model without resorting to the partially characterized molecular details of the directional sensing system. The phenomenological approach has allowed us to clearly analyze the relation between the symmetries in the equations and their behavioral consequences, i.e., temporal FCD, spatial FCD and spatio-temporal FCD. As future work, a more detailed model will be helpful to capture the exact molecular mechanism of the system. For example, we have adopted a nonlinear function $S^n/(S^n + (KB)^n)$ in Eqs. 1 without specifying how it is implemented at the molecular level. Such a sigmoidal nonlinearity is known to emerge based on various mechanisms such as cooperative interactions of signaling molecules [33], zero-order ultrasensitivity [34], and multistep effects [35], just to name a few. On the other hand, in the nonlinear function, the variable *B* works as a memory in the sense that *B* alters *A*'s sensitivity depending on the recent history of *S*. A similar input function can be seen in the model for bacterial chemotaxis too where the molecular mechanism for the memory has been identified as the methylation level of the chemoattractant receptors [36–38]. We discuss a biochemical network that enables such a sensitivity control in the Supporting Text S1.

In summary, we have combined two major concepts in quantitative biology, namely, FCD and the LEGI framework, and have proposed a new system-level mechanism to explain the rescaling behaviors of the gradient sensing system. We hope that our results promote deeper understanding of the directional sensing in eukaryotic cells.

# Author Contributions

K.K. and Y.K. designed research, performed research, analyzed data and wrote the paper. K.K. and Y.K contributed equally to this work.

# Acknowledgments

We thank S. Sawai and the members of the Sawai lab for discussions, M. Hamidi, and T. S. Shimizu for comments on the manuscript. This work is supported by The Paul G. Allen Family Foundation and the Foundation for Fundamental Research on Matter (FOM), which is part of the Netherlands Organisation for Scientific Research (NWO) (to K. K.) and the Platform Project for Supporting in Drug Discovery and Life Science Research (Platform for Dynamic Approaches to Living Systems) from Japan Agency for Medical Research and Development (AMED) and JSPS KAKENHI Grant Number 26840077. (to Y. K.).

# Supporting Text S1

Rescaling of spatio-temporal sensing in eukaryotic chemotaxis

Keita Kamino, [1][¶][*] Yohei Kondo, [2][¶][*]

[1]FOM Institute AMOLF, Amsterdam, Netherlands, [2]Graduated school of Informatics, Kyoto University, Kyoto, Japan,

*Corresponding author

[¶]These authors contributed equally to this work.

**Earlier models for the eukaryotic gradient sensing system and their symmetric properties**

In the main text, we have shown that, based on the LEGI-framework [1], spatial and temporal FCDs can be achieved respectively by two different scale-invariant properties in the governing equations, i.e., the spatial and temporal FCD symmetry (Fig. 5a). The two symmetries can be implemented in a mathematical model independently of each other (Fig. 5). The scale-invariant LEGI model (Eqs. 1) satisfies the two symmetries at the same time, and thus the model shows characteristic rescaled responses to stimuli that change in both space and time like traveling wave stimuli (Fig. 6). Among the two symmetric properties, the temporal FCD symmetry and the system-level consequence, i.e., temporal FCD, have been investigated extensively based on ordinary differential equation systems [2,3] and applied to directional sensing in spatially-homogeneous stimuli [4]. Concerning spatial FCD, although several LEGI-based models have the spatial FCD property as pointed out by Nakajima et al [5], the symmetric properties of the models, to our best knowledge, have not been studied explicitly. Here, we analyze representative LEGI-based models in the literature in terms of their symmetric properties.

(i) Levchenko-Iglesias models [6]

In the pioneering work by Levchenko and Iglesias [6], LEGI-based models have been proposed to describe the response properties of the gradient sensing system. They have assumed that the signaling component characterizing the response is found in both an active, $R^*$, and an inactive, $R$, states. The conversion from $R^*$ to $R$ is mediated by the inactivator $I$ and the opposite reaction is mediated by the activator $A$. Both of the activator $A$ and inactivator $I$ are activated by chemoattractant $S$. The first model equations they have proposed (Fig. 3A in [6]) are

$$\frac{\partial A}{\partial t} = -k_{-A}A + k_A S,$$

$$\frac{\partial I}{\partial t} = -k_{-I}I + k'_I S + D\nabla^2 I,$$

$$\frac{\partial R^*}{\partial t} = -k_{-2}IR^* + k_2 AR,$$

$$\frac{\partial R}{\partial t} = -k_{-1}IR + k_{1a}A + k_{-2}IR^* - k_2 AR. \quad (S1)$$

The rate constants are denoted as $k_{-A}$, $k_A$, $k_{-I}$, $k'_I$, $k_{-2}$, $k_2$, $k_{-1}$, $k_{1a}$, $k_{-2}$ and $k_2$. The diffusion coefficient of the inhibitor $I$ in the cytosolic region is represented by $D$. Note that, by the first and the second terms in the RHS of the fourth equation, the activator $A$ and the inactivator $I$ produce and degrade the inactive state of the response element $R$, respectively, which contribute to the amplification of the output signal $R^*$ [5,6]. The capability of the system to show spatial FCD in a static spatially-graded stimulus [5,6] can be understood by the spatial FCD symmetry of the model; at a steady state, i.e, $\frac{\partial A}{\partial t} = \frac{\partial I}{\partial t} = \frac{\partial R^*}{\partial t} = \frac{\partial R}{\partial t} = 0$, solutions of Eqs. S1 are invariant under a scale transformation

$$(A, I, R^*, R, S) \rightarrow (pA, pI, R^*, R, pS) \quad (p > 0). \quad (S2)$$

However, without assuming a steady state, the scale invariance does not hold, and thus the system does not show the temporal FCD property. Another model proposed by the authors combines the similar circuit to Eqs. S1 with an amplification module consisting of new variables, $T$ and $T_{in}$ (Fig. 3B in [6]):

$$\frac{\partial A}{\partial t} = -k_{-A}A + k_A S,$$

$$\frac{\partial I}{\partial t} = -k_{-I}I + k'_I S + D\nabla^2 I,$$

$$\frac{\partial R^*}{\partial t} = -k_{-R}IR^* + k_R A(R_{tot} - R^*),$$

$$\frac{\partial T}{\partial t} = k_T \frac{R^* T_{in}}{K_T + T_{in}} - k_\rho T,$$

$$\frac{\partial T_{in}}{\partial t} = -k_T \frac{R^* T_{in}}{K_T + T_{in}} + k_\rho T + \sigma + k_\sigma T - \gamma_\sigma T_{in}, \quad (S3)$$

where $k_{-A}$, $k_A$, $k_{-I}$, $k'_I$, $k_{-R}$, $k_R$, $k_T$, $k_\rho$, $k_\sigma$, $K_T$ and $\gamma_\sigma$ are rate constants. $D$ and $\sigma$ are the diffusion coefficient and the rate of constant supply of $T_{in}$, respectively. The total amount of the response elements ($R_{tot}$) is conserved. In addition to a feedforward circuit described by the first three equations, the fourth and fifth equations compose an amplification module based on a positive feedback loop, which locates at the downstream of the feedforward circuit [6]. The system (Eqs. S3) shows essentially the same behavior as the previous one (Eqs. S2) from the point of view of the response rescaling. Namely, the equations are scale invariant upon a scale transformation,

$$(A, I, R^*, T, T_{in}, S) \rightarrow (pA, pI, R^*, T, T_{in}, pS) \quad (p > 0) \quad (S4),$$

only when the system is in a steady state. Thus, the temporal FCD is not achieved. However, the system shows the temporal FCD if we assume the kinetics of the third equation is sufficiently fast. In this case, the third equation becomes an algebraic equation as

$$R^* = \frac{k_A A}{k_A A + k_{-R} I} R_{tot}. \quad (S5)$$

Then, the system is now invariant to the scale transformation (Eqs. S4) without assuming a steady state and therefore shows both spatial and temporal FCDs.

Following this scheme, Wang et al. [7] have proposed a model in which the upstream module is the same as the first three equations in Eqs. S3 while the downstream amplification module is replaced by another one. Hence, the same logic as in Eqs. S3 applies to their model. Namely, it satisfies the spatial FCD symmetry without modification and, by assuming fast kinetics of one variable, it can achieve both symmetries.

(ii) A balanced inactivation model [8]

Levine et al. have proposed a LEGI-based model which, instead of assuming downstream amplification module as in Eqs. S3, introduces a hypothetical mutual inhibition between chemical components to describe amplification and adaptation [8]. The governing equations are

$$\frac{\partial A}{\partial t} = k_a S - k_{-a} A - k_i A B_m \text{ at the membrane,}$$

$$\frac{\partial B_m}{\partial t} = k_b B - k_{-b} B_m - k_i A B_m \text{ at the membrane, and}$$

$$\frac{\partial B}{\partial t} = D \nabla^2 B \text{ in the cytosol} \quad (S6),$$

with a boundary condition

$$D \frac{\partial B}{\partial n} = k_a S - k_b B,$$

where the derivative is the ourward pointing normal derivative of the cytosolic component $B$. The rate constants are represented by $k_a$, $k_{-a}$, $k_i$, $k_b$ and $k_{-b}$, respectively. $D$ is the diffusion coefficient of the cytosolic component $B$. Due to the mutual inhibition term ($k_i A B_m$) in the first and second equations, the system does not show any scale invariance upon scale change in the input stimulus, i.e., $S \to pS$. As a result, the system shows neither temporal FCD (Fig. 1 in [8]) nor spatial FCD. In fact, the output level $A$ in a steady state does not show perfect adaptation and depends on the absolute level of the input stimulus $S$ as shown analytically in [8].

(iii) Ultrasensitive LEGI model [5]

An alternative LEGI-based model has been proposed by Nakajima et al [5] to describe the responses in traveling wave stimuli. The governing equations are

$$\frac{\partial A}{\partial t} = k_a S - \gamma_a A,$$

$$\frac{\partial I}{\partial t} = k_i S - \gamma_i I + D \nabla^2 I,$$

$$\frac{\partial R}{\partial t} = Ak_A \frac{R_{tot} - R}{K_A + (R_{tot} - R)} - Ik_I \frac{R}{K_I + R} \quad (S7),$$

where $k_a$, $\gamma_a$, $k_i$, $\gamma_i$, $k_A$, $K_A$, $k_I$ and $K_I$ are rate constants. $D$ is the diffusion coefficient of the inhibitor $I$. This model has essentially the same upstream module as the model (S3), but the output signal $R$ is now regulated by the functions of Michaelis-Menten form, which is critical for the highly asymmetric response to spatially-symmetric traveling wave stimuli [5]. In the same way as the previous model (Eqs. S3), this model at a steady state satisfies the spatial FCD symmetry under the transformation,

$$(A, I, R, S) \rightarrow (pA, pI, R, pS) \quad (p > 0) \quad (S8).$$

This explains the rescaled response to a gradient stimulus of the model [5]. Also in the same way as the previous model (Eqs. S3), by assuming fast kinetics of the third equation, the equations satisfie the temporal FCD symmetry.

(iv) LEGI-BEN model [9]

Tang et al. [9] have recently proposed a model based on the so-called LEGI-BEN scheme where a LEGI module is combined with a downstream excitable system [10]. Although the primary interest of the line of research is to connect gradient sensing with migration, here we focus only on the upstream LEGI module and its rescaling property. The model equations are

$$\frac{\partial A}{\partial t} = k_a S - \gamma_a A,$$

$$\frac{\partial I}{\partial t} = k_i S - \gamma_i I + D \nabla^2 I,$$

$$\frac{\partial R}{\partial t} = k_R \frac{k_A + A}{k_I + I} - \gamma_R R \quad (S9),$$

where $k_a$, $\gamma_a$, $k_i$, $\gamma_i$, $k_R$, $k_A$, $k_I$ and $D$ are constant parameters. The model shows neither the spatial nor temporal FCD symmetry on its own. However, the model can show spatio-temporal FCD under a condition without assuming fast kinetics of any variable like we have done in the previous sections. A characteristic feature of the model is that the inhibitor $I$ down-regulates the output $R$ by suppressing the synthesis of $R$ by the activator $A$, which is described by the first term on the right hand side of the third equation. Because of the term, in the limit of $k_A, k_I \rightarrow 0$ (or the limit of large input), the model becomes to satisfy both space and temporal FCD symmetry under the transformation,

$$(A, I, R, S) \rightarrow (pA, pI, R, pS) \quad (p > 0) \quad (S10).$$

In the original LEGI-BEN model, the rescaled output from the upstream module is fed into the downstream system (i.e., the excitable system) and therefore assures rescaled behaviors of the entire system.

**A possible molecular mechanism for the nonlinear activation function**

In the scale-invariant LEGI model (Eqs. 1 in the main text), we adopted a nonlinear function $S^n/(S^n + (KB)^n)$ in which the inhibitor $B$ suppresses the activator $A$ by lowering the sensitivity to the input $S$, not by degrading the activator $A$ directly. Below, we show how the sensitivity control can emerge in a biochemical network. Let us introduce a path inhibition model as follows:

$$\frac{\partial M}{\partial t} = k_m S - k_{-m} B M$$
$$\frac{\partial A}{\partial t} = k_a \frac{M^n}{M^n + K^n} - k_{-a} A,$$
$$\frac{\partial B}{\partial t} = k_b S - k_{-b} B + D(\langle B \rangle - B) \quad (S11),$$

where $k_m$ and $k_{-m}$ represent the rate constants of the equation for $M$ and the other parameters and variables are the same as in Eqs. 1. In this model, a signaling component $M$ mediates excitatory regulation from the input $S$ to the activator $A$, and the inhibitor $B$ suppresses the mediator $M$ instead of the activator signal directly (hence it is called a 'path inhibition' model). When the kinetics of $M$ is fast, the mediator level $M$ is approximated by $S/B$ (Here we assume $k_m/k_{-m} = 1$ for simplicity). By substituting $M = S/B$ into the equation for $A$, we obtain the nonlinear sensitivity control function. The resultant model is equivalent to the scale-invariant LEGI model (Eqs. 1).

gradients. Curr Opin Cell Biol. Elsevier Ltd; 2012;24: 245–253. doi:10.1016/j.ceb.2011.11.009

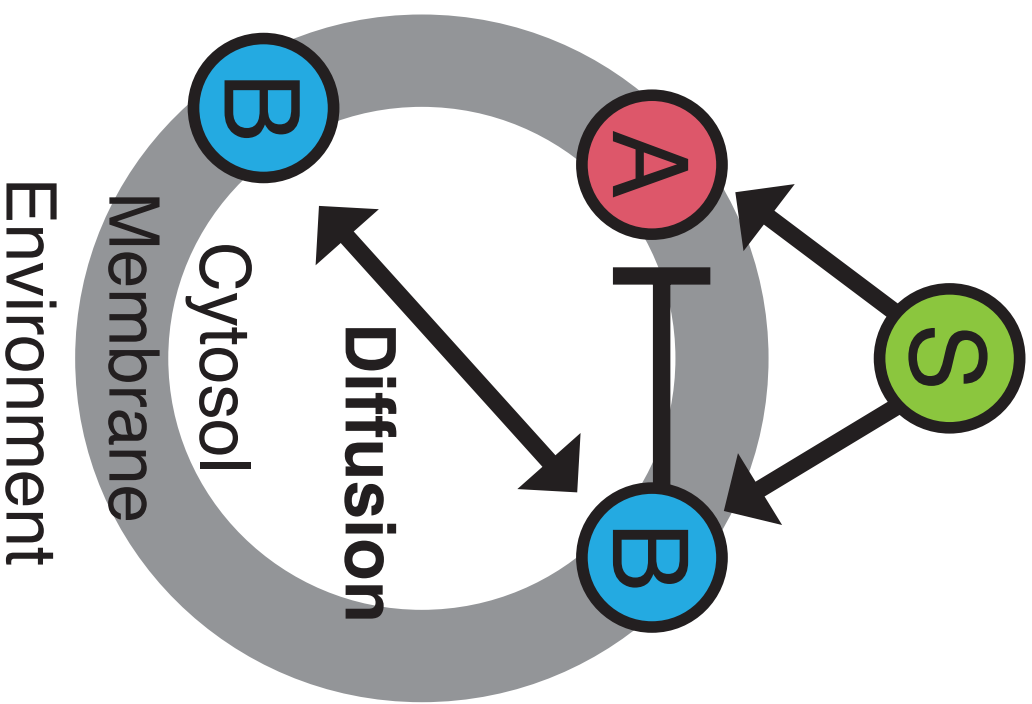

Fig. 1

Fig. 2

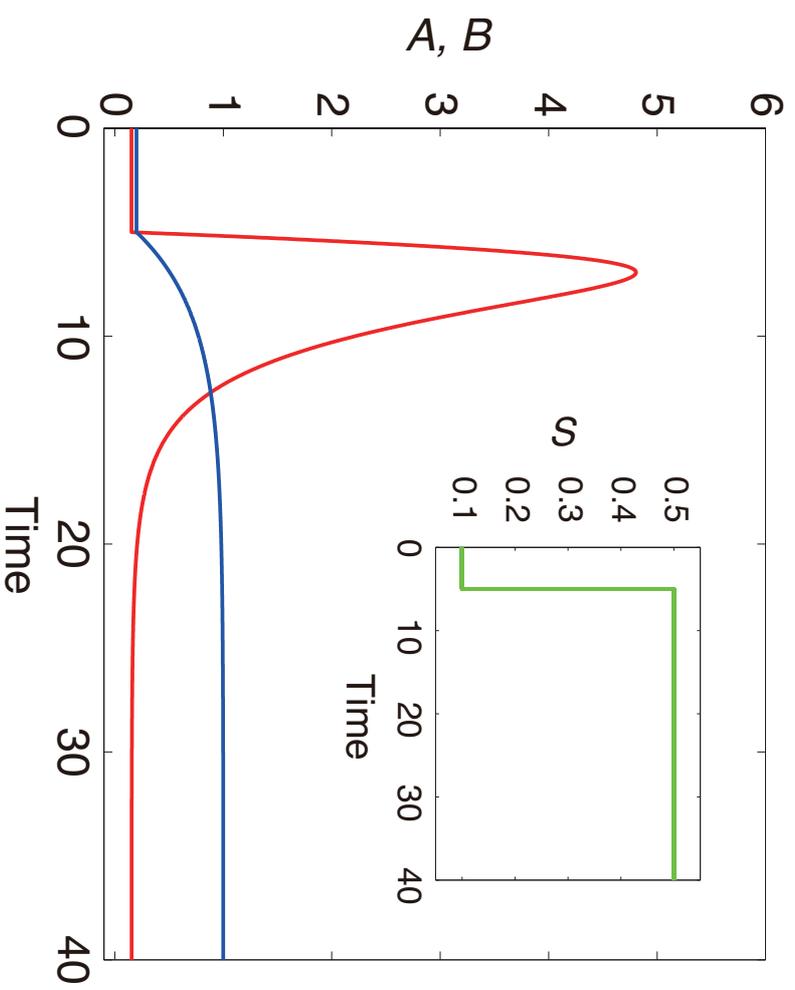

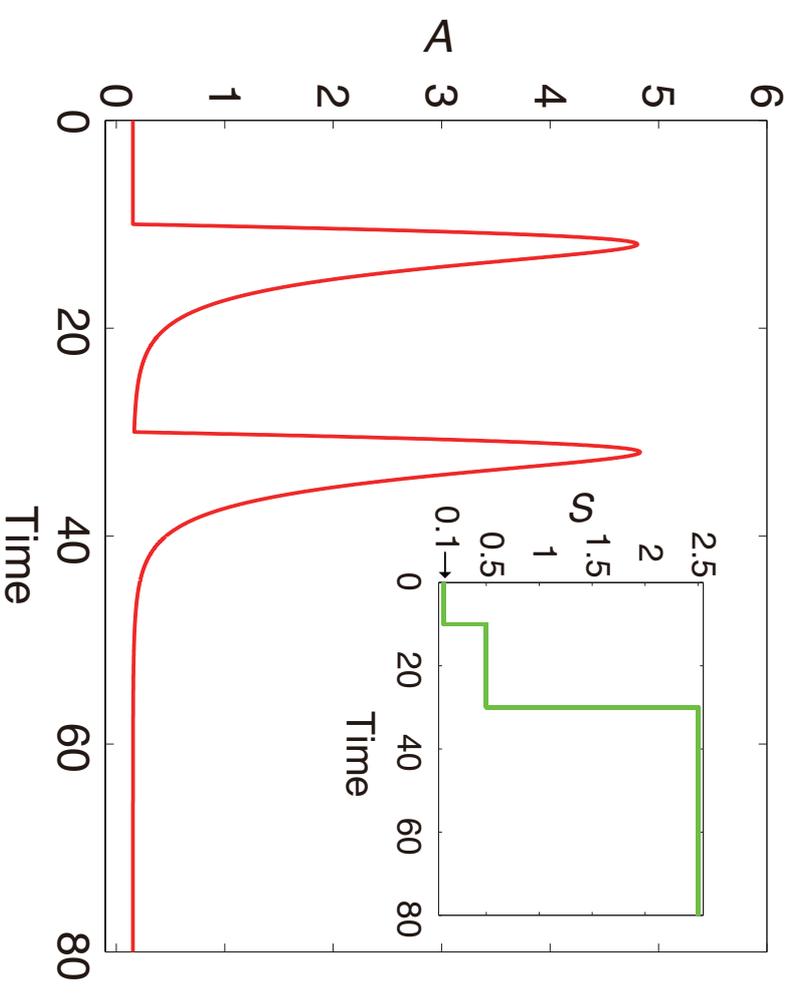

Fig. 3

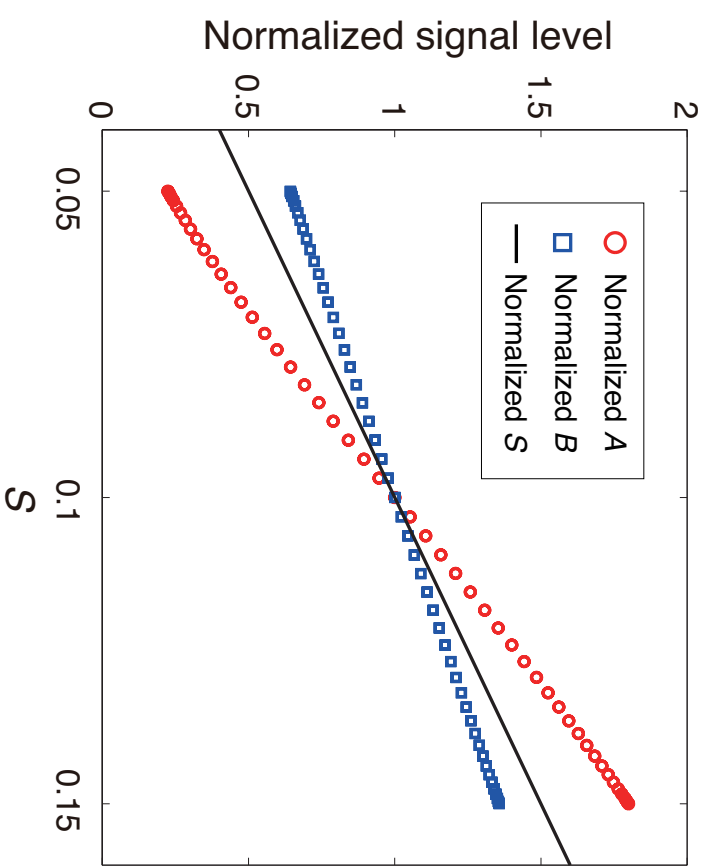
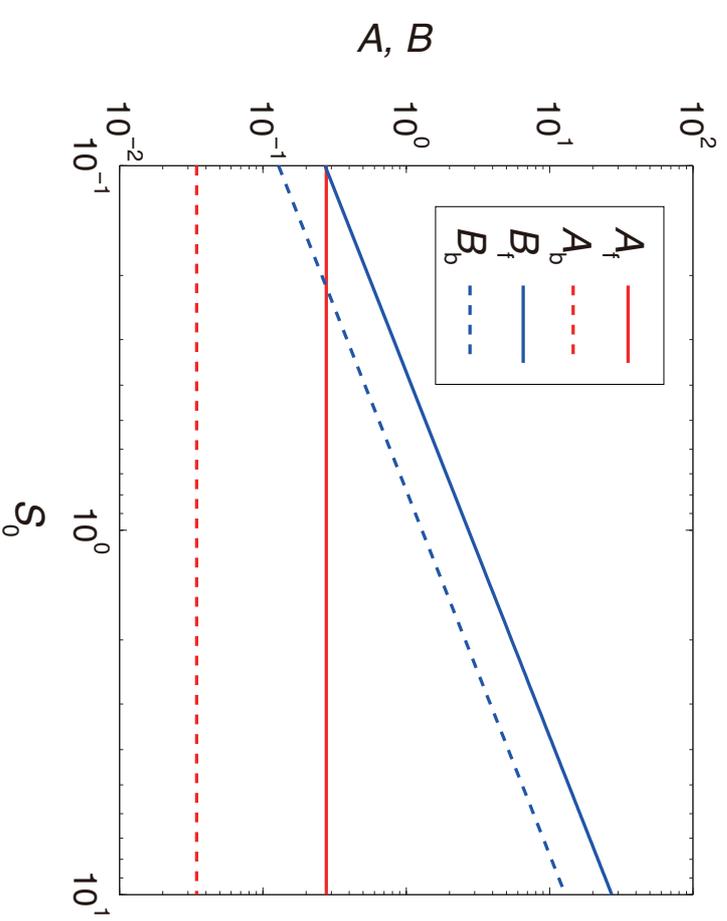

Fig. 4

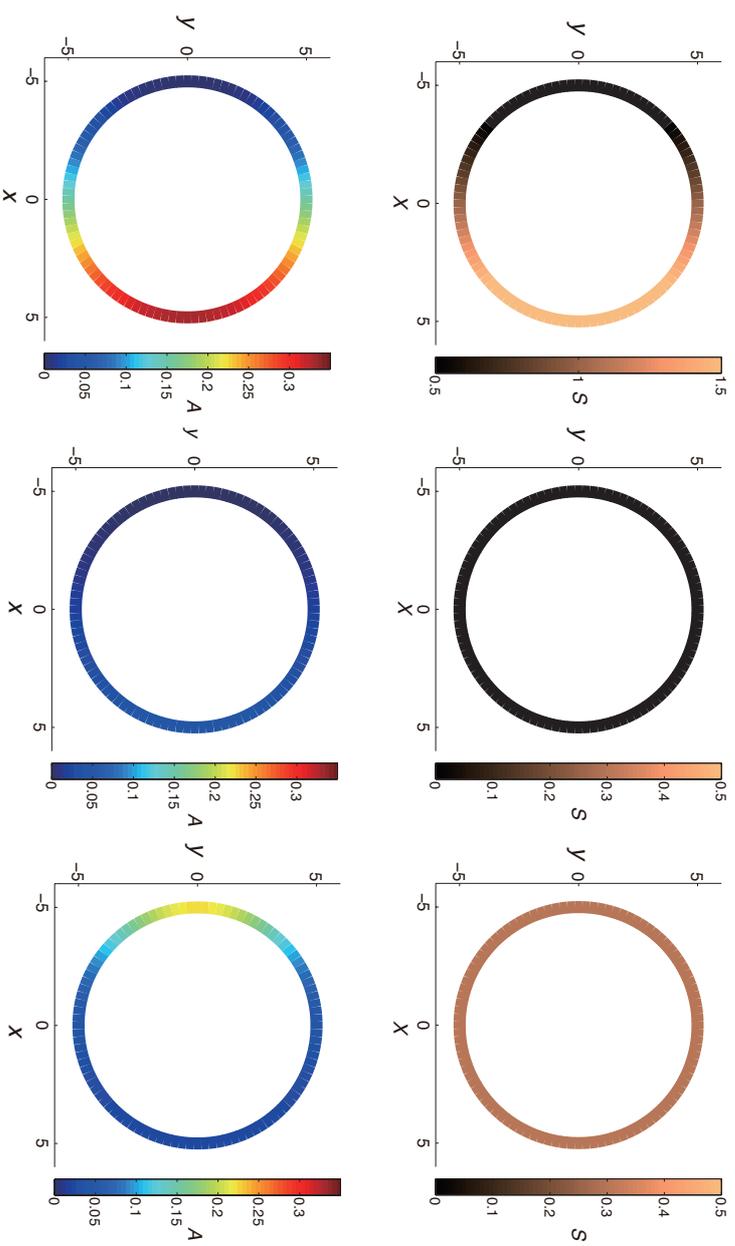
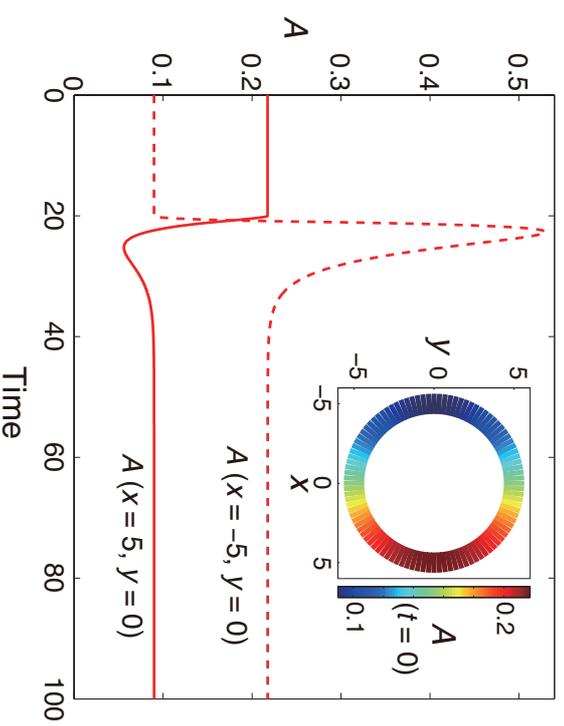
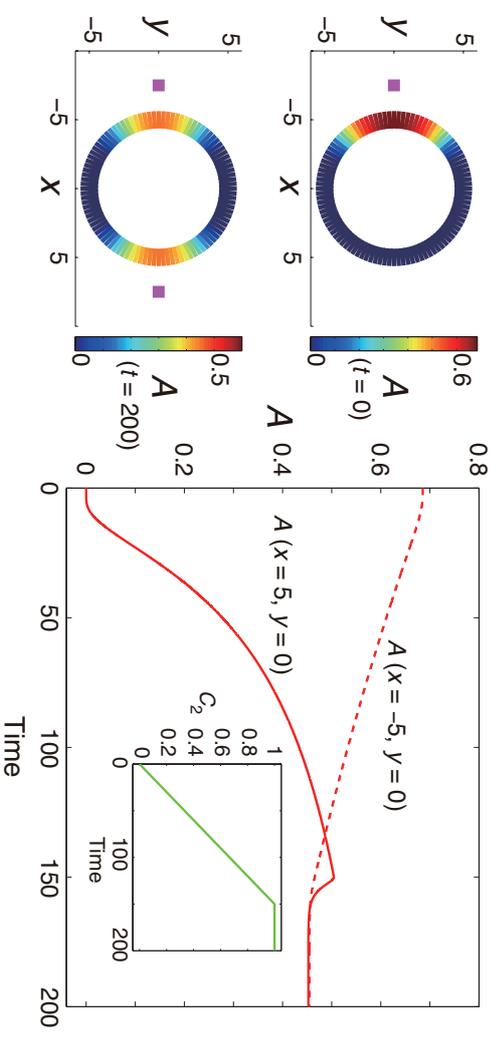

# Fig. 5

**a**

Temporal FCD symmetry
$$\frac{\partial A}{\partial t} = \text{reaction terms}$$

Spatial FCD symmetry
$$\frac{\partial B}{\partial t} = \text{reaction terms} + \text{coupling terms}$$

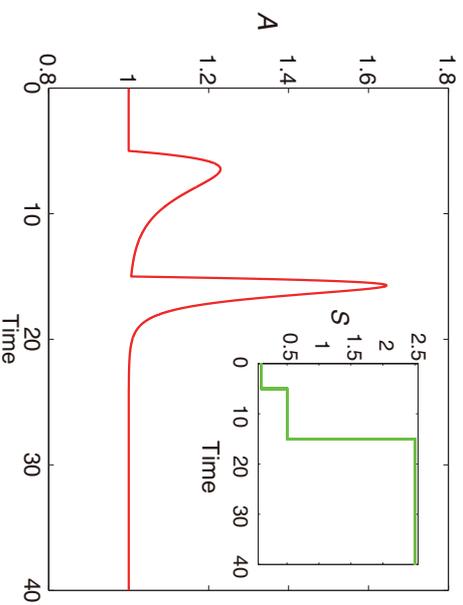

**b**

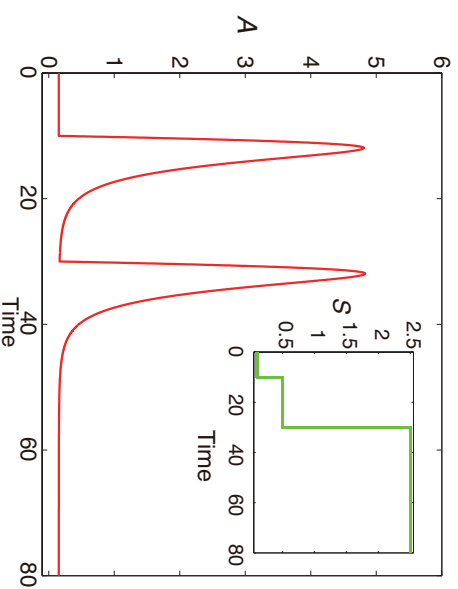

**c**

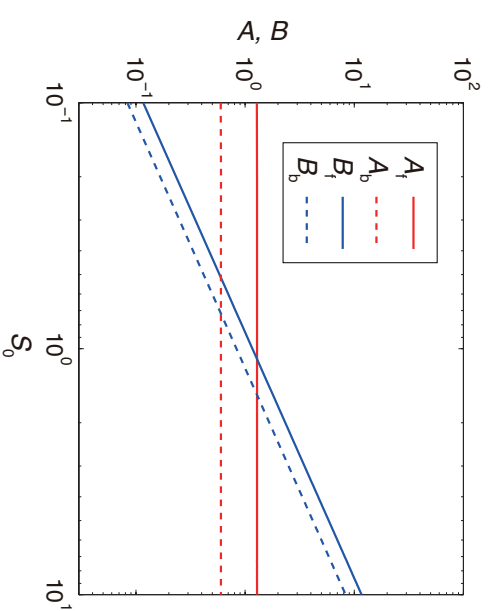

**d**

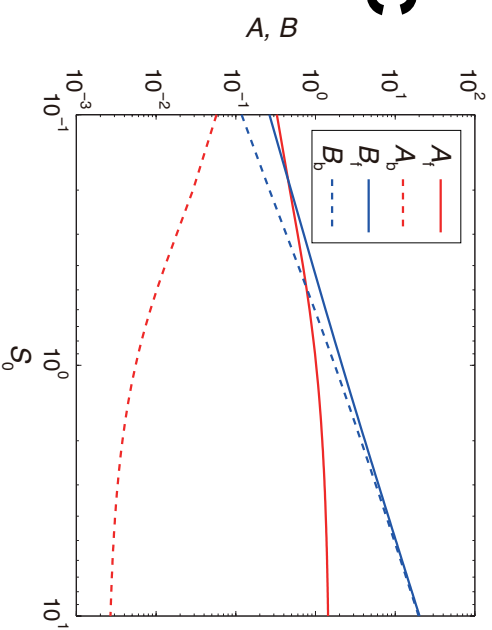

**e**

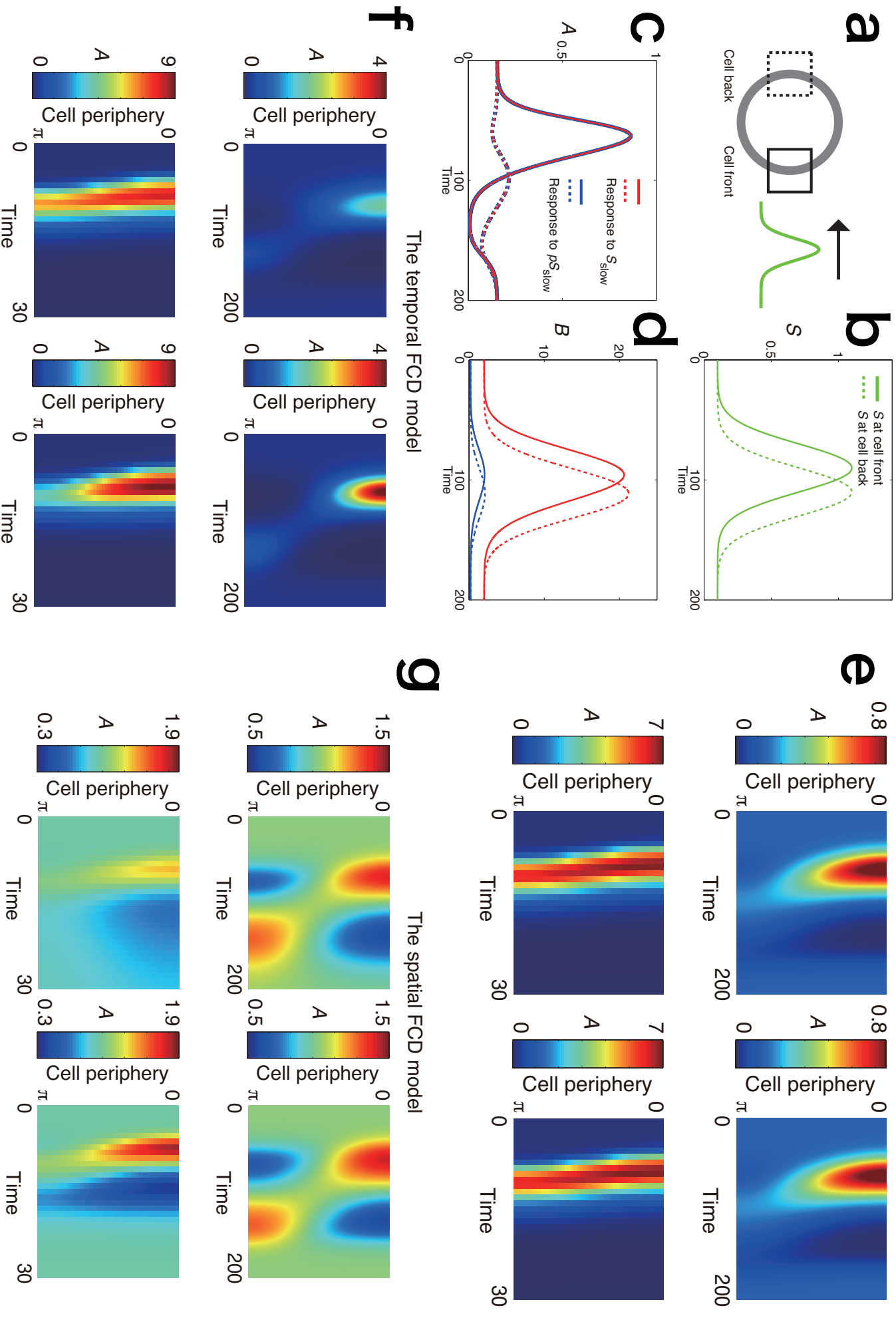

Fig. 6

# Fig. 7

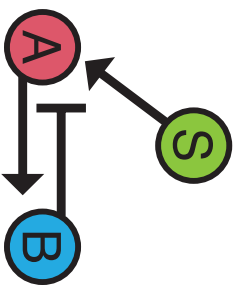

a

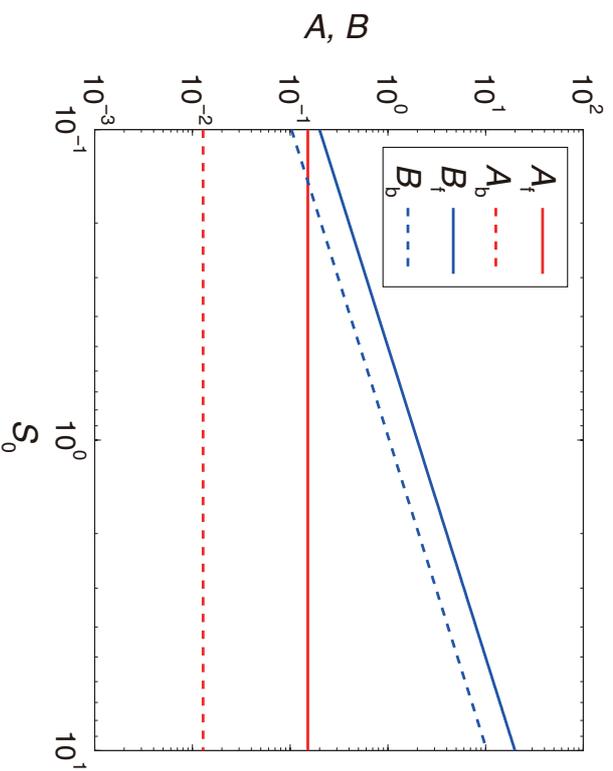

b

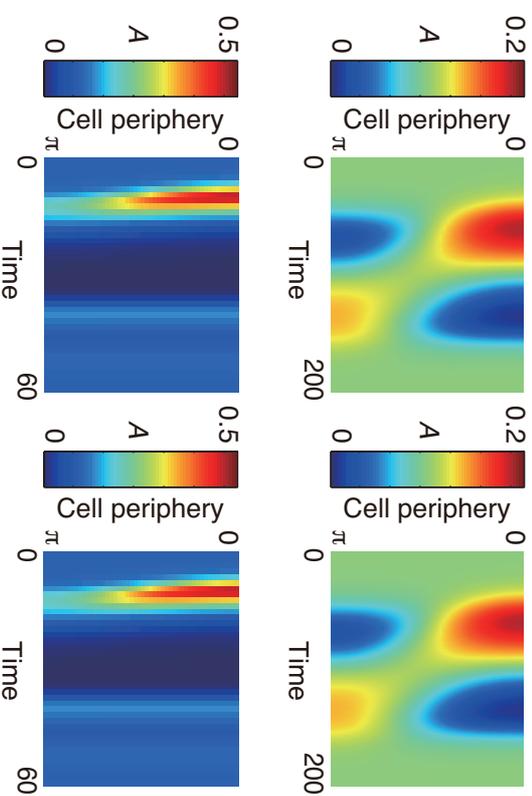

c

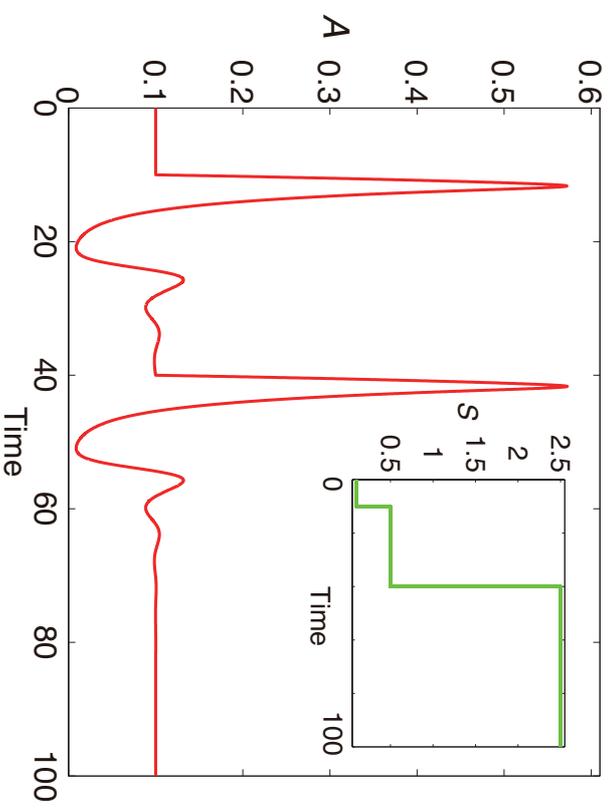

d